\documentclass[reprint,amsmath,amssymb,aps,floatfix]{revtex4-2}
\usepackage{multirow}
\usepackage{graphicx}
\usepackage{dcolumn}
\usepackage{bm}
\usepackage[mathlines]{lineno}

\usepackage[showframe],

\usepackage{siunitx}
\usepackage{xcolor}

\begin{document}

\preprint{APS/123-QED}

\title{Magnetic Ordering in GdAuAl$_4$Ge$_2$ and TbAuAl$_4$Ge$_2$: layered compounds with triangular lanthanide nets} 

\author{Keke Feng}
\altaffiliation {Department of Physics, Florida State University.}
\altaffiliation {National High Magnetic Field Laboratory}
\author{Ian Andreas Leahy}
\altaffiliation {Department of Physics, University of Colorod Boulder}
\author{Olatunde Oladehin}
\altaffiliation {Department of Physics, Florida State University.}
\altaffiliation {National High Magnetic Field Laboratory}
\author{Kaya Wei}
\altaffiliation {National High Magnetic Field Laboratory}
\author{Minhyea Lee}
\altaffiliation {Department of Physics, University of Colorod Boulder}
\author{Ryan Baumbach}
\altaffiliation {Department of Physics, Florida State University.}
\altaffiliation {National High Magnetic Field Laboratory}
\date{\today}

\begin{abstract}
We report the synthesis of the entire $Ln$AuAl$_4$Ge$_2$ ($Ln$ = Y, Pr, Nd, Sm, Gd, Tb, Dy, Ho, Er, and Tm) series and focus on the magnetic properties of GdAuAl$_4$Ge$_2$ and TbAuAl$_4$Ge$_2$. Temperature and magnetic field dependent magnetization, heat capacity, and electrical resistivity measurements reveal that both compounds exhibit several magnetically ordered states at low temperatures, with evidence for magnetic fluctuations extending into the paramagnetic temperature region. For magnetic fields applied in the $ab$-plane there are several ordered state regions that are associated with metamagnetic phase transitions, consistent with there being multiple nearly degenerate ground states. Despite Gd being an isotropic $S$-state ion and Tb having an anisotropic $J$-state, there are similarities in the phase diagrams for the two compounds, suggesting that factors such as the symmetry of the crystalline lattice, which features well separated triangular planes of lanthanide ions, or the Ruderman-Kittel-Kasuya-Yosida interaction as defined by the Fermi surface topography control the magnetism. We also point out similarities to other centrosymmetric compounds that host skyrmion lattices such as Gd$_2$PdSi$_3$, and propose that the $Ln$AuAl$_4$Ge$_2$ family of compounds are of interest as reservoirs for complex magnetism and electronic behaviors such as the topological Hall effect.

\end{abstract}
\pacs{Valid PACS appear here}       
\maketitle

\section{\label{sec:intro}Introduction}

There is ongoing interest in magnetic skyrmions because (i) they may be useful for spintronics applications or as new generation of information storage devices and (ii) they present opportunities for basic science~\cite{Kim_2019_review,skyrmion-review}. Early work focused on materials such as MnSi~\cite{MnSi} and Fe$_{\rm{0.5}}$Co$_{0.5}$Ge~\cite{CoGe}, where the non-centrosymmetric crystal structure produces a Dzyaloshinskii-Moriya (DM) interaction. More recently, there has been a surge of interest in centrosymmetric lanthanide metals that exhibit topologically non-trivial magnetic textures in the absence of a DM interaction~\cite{Skyrmion.centrosymmetric.condition.Nature.Communication.2015,skyrmion.Centrosymmetric.Magnets.PRL.2018,Gd3Ru4Al12.skyromion.Nature.Communication.2019,GdRu2Si2.Nature.2020,Gd2PdSi3.skyrmion.science.2019,TmB4.PRB.2016,EuAl42021}. In these materials, a delicate balance between the crystalline anisotropy and various magnetic energy scales (e.g., geometric frustration~\cite{Kagome.PRB.2009,Pyrochlore.Lattice.PRL.2018}, competing RKKY interactions~\cite{ANNNI.model.1988,CeSb.1977.PRB,KuanWen.PRB.2017}), and crystal electric field effects~\cite{PRL.CEF,PRB.CEF.2017,PRB.CEF.2001}) combine to produce their complex magnetic states.
It is also seen that these materials feature unexpected coupling between different degrees of freedom, where a hallmark behavior is the topological Hall effect~\cite{THE1,THE2}. 

Motivated by this, we recently investigated the weakly correlated $f$-electron metal CeAuAl$_4$Ge$_2$~\cite{1142,Shengzhi2017}, where the planar triangular arrangement of the cerium ions resembles what is seen for Gd$_2$PdSi$_3$~\cite{Gd2PdSi3.skyrmion.science.2019,saha99}. In this case, we demonstrated that the cerium ions are trivalent, are weakly interacting, and there is limited evidence for magnetic frustration. This led us to consider whether strengthened magnetic interactions could drive magnetic frustration: e.g., by replacing Ce with other lanthanide ($Ln$) ions. Here we report the synthesis of single crystals of the entire $Ln$AuAl$_4$Ge$_2$ series and focus on the magnetic properties of GdAuAl$_4$Ge$_2$ and TbAuAl$_4$Ge$_2$. These examples not only are expected to have have large effective magnetic moments ($\mu_{\rm{eff}}$ $=$ 7.94 $\mu_{\rm{B}}$/Gd and 9.72 $\mu_{\rm{B}}$/Tb), but also allow a comparison between a pure spin ion (Gd$^{3+}$; $S$ $=$ 7/2, $L$ $=$ 0, $J$ $=$ 7/2) and a mixed spin-orbital ion (Tb$^{3+}$;  $S$ $=$ 3, $L$ $=$ 3, $J$ $=$ 6). Temperature and field dependent magnetization measurements reveal that GdAuAl$_4$Ge$_2$ shows little anisotropy in the paramagnetic state while TbAuAl$_4$Ge$_2$ shows easy $ab$-plane anisotropy. Complex magnetic ordering appears at low temperatures for both compounds, where GdAuAl$_4$Ge$_2$ has three magnetic phase transitions at $T_{\rm{N1}}$ = 17.8 K, $T_{\rm{N2}}$ = 15.6 K, and $T_{\rm{N3}}$ = 13.8 K and TbAuAl$_4$Ge$_2$ exhibits two magnetic phase transitions at $T_{\rm{N1}}$ = 13.9 K and $T_{\rm{N2}}$ = 9.8 K. All of these transitions are antiferromagnetic-like in low fields, but magnetic fields applied in the $ab$ plane drive metamagnetic phase transitions into spin polarized states. 

From these measurements, we construct $T-H$ magnetic phase diagrams, where there are several regions with complex magnetic configurations. Heat capacity measurements additionally show that a substantial fraction of the magnetic entropy is found at temperatures above the ordered states, consistent with the presence of magnetic fluctuations (and crystal electric field effects for TbAuAl$_4$Ge$_2$)~\cite{Kumar_2008,GdHC,Mallik_1998,HC-frust}. Finally, the temperature dependent electrical resistivity for GdAuAl$_4$Ge$_2$ exhibits a broad minimum that precedes the ordered state, providing evidence for possible magnetic frustration~\cite{resmin1,Kumar_2020}. Taken together, these measurements reveal that although these compounds form with a relatively simple crystal structure, they nonetheless exhibit magnetic degeneracy that resembles what is seen for other centrosymmetric skyrmion lattices such as Gd$_2$PdSi$_3$. This invites further investigation to determine whether these compounds are hosts for magnetically ordered states with multiple q-vectors or nontrivial magnetic topologies, where phenomena such as the topological Hall effect might be observed.


\section{\label{sec:methods}Experimental Methods}

$Ln$AuAl$_4$Ge$_2$ ($Ln$ = Y, Pr, Nd, Sm, Gd, Tb, Dy, Ho, Er, and Tm) single crystals were grown using an aluminum molten metal flux, as previously reported~\cite{1142}. Elements with purities $>$ 99.9$\%$ were combined in the molar ratio (RE):1(Au):10(Al):5(Ge) and loaded into 2-mL alumina Canfield crucibles~\cite{PCCanfield}. The crucibles were sealed under vacuum in quartz tubes, heated to 1000 $^{\circ}$C at a rate of 83$^{\circ}$C/h, kept at 1000$^{\circ}$C for 15 h, then cooled to 860$^{\circ}$C at a rate of 7$^{\circ}$C/h and held at 860 $^{\circ}$C for 48 h. This was followed by cooling down to 700$^{\circ}$C at a rate of 12$^{\circ}$C/h. Excess flux was removed by centrifuging the tubes at 700$^{\circ}$C, after which single-crystal platelets were collected. Crystals exhibited triangular facets that are associated with the $ab$ plane.

Room temperature powder x-ray diffraction (PXRD) measurements were performed using a Rigaku SmartLab SE X-ray diffractometer with a Cu K$\alpha$ source. Crystal structure refinement analysis was performed using the Winprep software. The principal axes were identified using Enraf-Nonius CAD-4 diffractometer, where the orientation of the full reciprocal lattice is obtained. Magnetization $M$ measurements were carried out at temperatures $T$ = 1.8 - 300 K under applied magnetic fields of $\mu_0H$ = 0.5 - 9 T applied parallel ($\|$) to the crystallographic $c$ axis and the $ab$ plane using a Quantum Design VSM Magnetic Property Measurement System. Specific heat $C$ measurements were performed for temperatures $T$= 1.8 - 70 K in a Quantum Design Physical Properties Measurement Systems using a conventional thermal relaxation technique. DC electrical resistivity $\rho$ measurements for temperatures $T$= 1.8 - 300 K were performed in a four-wire configuration for polished single crystal using the same system.


\section{\label{sec:Results}Results}

\subsection{$Ln$AuAl$_4$Ge$_2$ X-ray Diffraction}
\begin{figure*}
  \includegraphics[width=\textwidth]{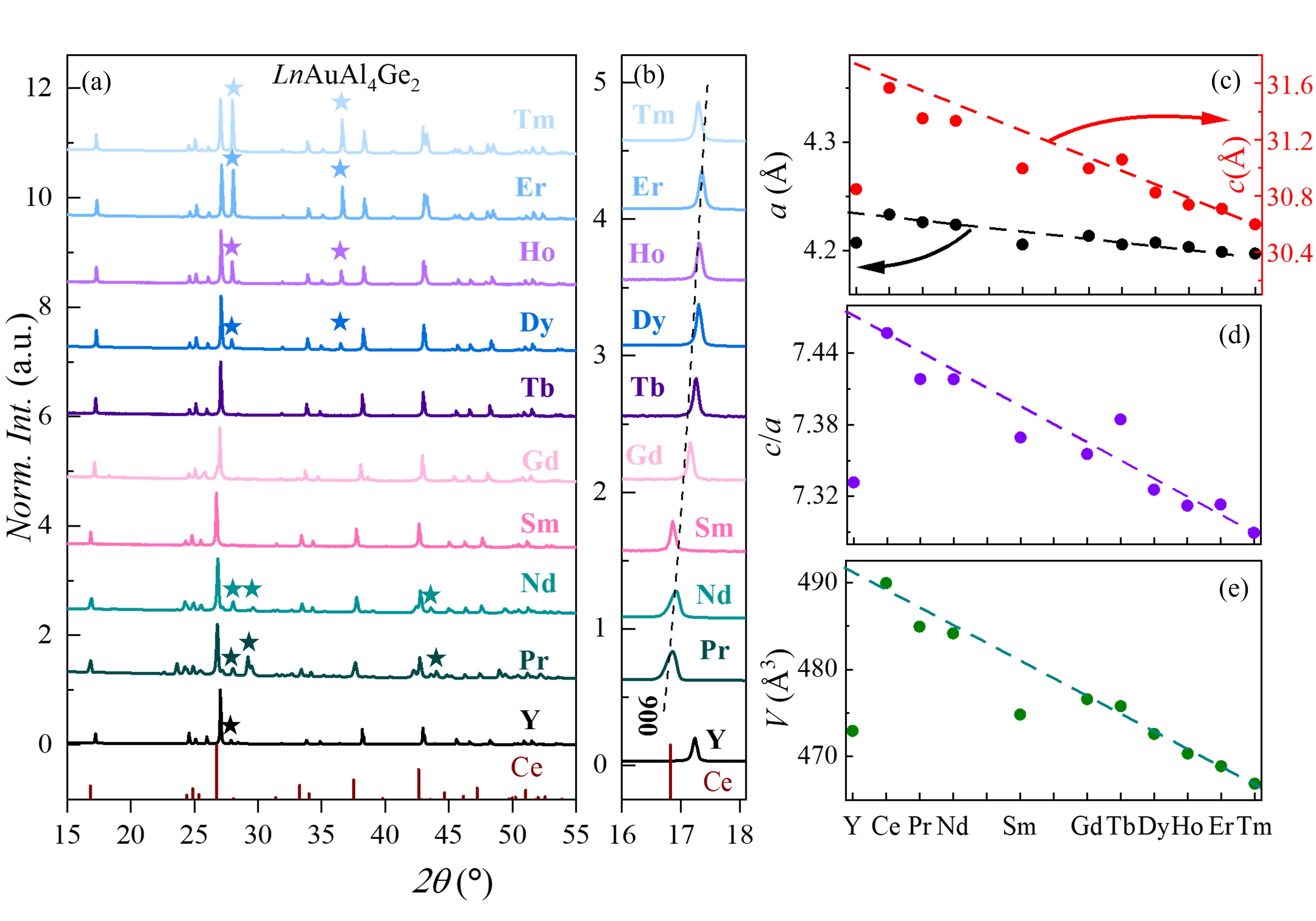}\caption{\label{fig1}(a) Powder x-ray diffraction data for $Ln$AuAl$_4$Ge$_2$ ($Ln$ = Y, Pr, Nd, Gd, Tb, Dy, Ho, Er, and Tm). The normalized intensity patterns are compared to the theoretical peaks for CeAuAl$_4$Ge$_2$~\cite{1142} and have been shifted vertically for clarity.  The impurity peaks are marked with stars. (b) $Ln$ dependence of the diffraction peak at 006. The dashed line is a guide to the eye. (c) The lattice constants $a$ and $c$ determined from Reitveld analysis of powder x-ray diffraction measurements. Results for Ce are from Ref.~\cite{1142}. (d) The ratio $c/a$ vs. $Ln$. (e) The unit cell volume $V$ vs. $Ln$. The dashed lines are guides to the eye.}
\end{figure*}

It was previously reported that CeAuAl$_4$Ge$_2$ crystallizes in a well-ordered rhombohedral structure with space group $R\overline{3}m$~\cite{1142,Shengzhi2017} where the cerium ions are in the trivalent state. In Fig.~\ref{fig1} we present powder x-ray diffraction patterns for the other members of the $Ln$AuAl$_4$Ge$_2$ series ($Ln$ = Y, Pr, Nd, Sm, Gd, Tb, Dy, Ho, Er, and Tm), showing that the same structure persists across the lanthanide series (except Eu). Production of phase pure specimens is challenging, as evidenced by the presence of extra peaks in the XRD pattern that do not belong to the $Ln$AuAl$_4$Ge$_2$ structure. For $Ln$ = Pr and Nd we were unable to index the additional peaks, but for $Ln$ = Dy, Ho, Er, and Tm they are consistent with the $Ln$Al$_2$Ge$_2$ structure~\cite{LnAl2Ge2.2002, ErAl2Ge2.2008}. In the latter case, there is a close similarity between $Ln$Al$_2$Ge$_2$ and the $Ln$AuAl$_4$Ge$_2$ phases, which likely relates to its tendency to appear in these crystals. We also find that the $Ln$ = Y, Ce, Gd, Tb, and Sm compound can be obtained as phase-pure crystals. Rietveld refinements were performed for the entire series and the obtained lattice parameters ($a$ and $c$), the ratio ($c$/$a$), and the unit cell volume ($V$) are plotted in Figs.~\ref{fig1}c-e, where their variation is consistent with a trivalent lanthanide contraction that isotropically compresses the unit cell. This behavior is expected since the radius of the lanthanide ions decreases with increasing atomic number, as long as the valence is fixed.

\begin{table*}
\begin{tabular}{|c|c|c|c|c|c|c|c|c|c| }
\hline
& $T_{\rm{N1}}$(K) &$T_{\rm{N2}}$(K) &$T_{\rm{N3}}$(K) &$H_{\rm{C1}}$(T) &$H_{\rm{C2}}$(T) &$H_{\rm{C3}}$(T) & $\theta$(K) & $\mu_{\rm{eff}}$($\mu_B$/F.U.)&$M_{\rm{sat}}$($\mu_B$/F.U.)  \\ 
\hline
 $Gd$& 17.8 &15.6 &13.8 &1.9 &--- &--- &2.3 &7.93&--- \\
\hline
 $Tb$& 13.9 &9.8  &--- &1.3 &1.9 &2.7 &18 &9.89&8.31 \\
 \hline
\end{tabular}
\caption{\label{tab:1} Summary of magnetic properties for GdAuAl$_4$Ge$_2$ and TbAuAl$_4$Ge$_2$ obtained from the magnetic susceptibility $\chi$($T$), the magnetization $M(H)$, and the heat capacity $C(T)$. $\chi(T)$ was collected in a magnetic field $\mu_0H$ $=$ 0.5 T. $C(T)$ and $\rho(T)$ were collected in $\mu_0H$ $=$ 0 T. $T_{\rm{N1}}$, $T_{\rm{N2}}$, and $T_{\rm{N3}}$ are the ordering temperatures, $H_{\rm{C1}}$, $H_{\rm{C2}}$, and $H_{\rm{C3}}$ are critical fields at $T$ $=$ 1.8 K, $\theta$ and $\mu_{\rm{eff}}$ are the Curie-Weiss temperatures and the effective magnetic moments obtained from fits to $\chi(T)$, and $M_{\rm{sat}}$ is that saturation moment obtained from $M(H)$.}
\end{table*}

\subsection{GdAuAl$_4$Ge$_2$}
\begin{figure}
\includegraphics[width=\columnwidth]{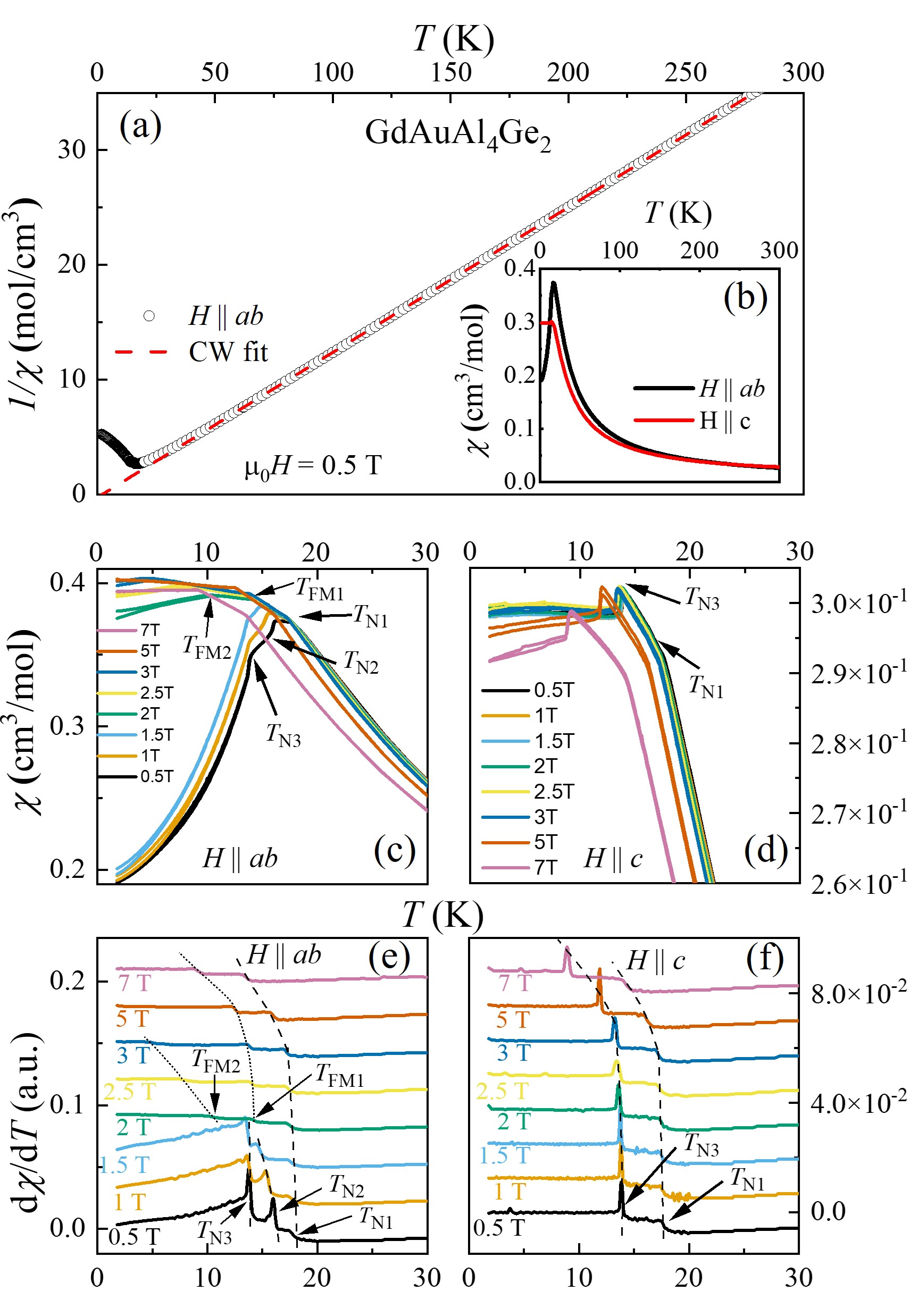}\caption{\label{fig2}(a) The inverse magnetic susceptibility $\chi^{-1} = (M/H)^{-1}$ vs temperature $T$ for GdAuAl$_4$Ge$_2$ collected in a magnetic field $\mu_0H$ $=$ 0.5 T applied parallel to $ab$-plane. The dotted line is a fit to the data using the Curie-Weiss law, as described in the text. (b) $\chi (T)$ for $\mu_0H$ $=$ 0.5 T applied both parallel (black line) to the $ab$-plane and c-axis (red line). (c,d) $\chi (T)$ for 1.8 K $<T<$ 30 K emphasizing the region near the magnetic ordering temperatures $T_{\rm{N1}}$, $T_{\rm{N2}}$, $T_{\rm{N3}}$, $T_{\rm{FM1}}$, and $T_{\rm{FM}}$ which are defined in the text. (e,f) Derivatives of the magnetic susceptibilities with respect to temperature $\frac{\partial \chi}{\partial T}$. Data have been shifted vertically for clarity, the dashed lines and short dot lines are guided to the eye.}
\end{figure}

The temperature-dependent magnetic susceptibility $\chi(T)$ $=$ $M/H$ for GdAuAl$_4$Ge$_2$ is shown in Fig.~\ref{fig2}. Curie-Weiss behavior with little anisotropy is seen for $T$ $=$ 50 - 300 K, and fits to the data using the expression $\chi$ = $C$/($T$-$\theta$) yield the parameters $\theta$ = 2.3 K and $\mu_{\rm{eff}}$ = 7.93 $\mu_B$/Gd ($\mu_{\rm{eff}}$ = 7.94 $\mu_B$ for Gd$^{3+}$). Anisotropic and field dependent magnetic ordering is seen at low temperatures (Fig.~\ref{fig2}c,d). For $\mu_0H$ $=$ 0.5 T applied in the $ab$-plane, there are three distinct transitions at $T_{\rm{N1}}$ = 17.8 K, $T_{\rm{N2}}$ = 15.6 K, and $T_{\rm{N3}}$ = 13.8 K which are defined from the derivative of the magnetic susceptibily $\partial \chi$/$\partial T$ (panels e and f). In the low field region, each transition decreases the magnitude of $\chi$ (i.e., they are antiferromagnetic-like) and they are suppressed with increasing $H$. There are abrupt changes in these trends above $\mu_0H$ = 1.5 T, where $\chi(T)$ no longer strongly decreases at $T_{\rm{N1}}$ and instead tends to saturate at low temperatures. This is consistent the occurrence of field driven metamagnetic phase transitions where the spins abruptly become more fully polarized. Within the spin polarized state, $T_{\rm{N1}}$ continues to be suppressed by field and there are additional features at $T_{\rm{FM1}}$ and $T_{\rm{FM2}}$ that suggest further weak spin reorientations. These transitions are seen as subtle features in $\chi(T)$ and its temperature derivative, but future work is needed to fully clarify the spin states that are associated with them. It is also seen that for $H$ $\parallel$ $c$, $\chi(T)$ only clearly shows the transitions at $T_{\rm{N1}}$ and $T_{\rm{N3}}$, which are gradually suppressed with $H$. The sharpness of the feature at $T_{\rm{N3}}$ suggests that it is first order.

\begin{figure}
  \includegraphics[width=\columnwidth]{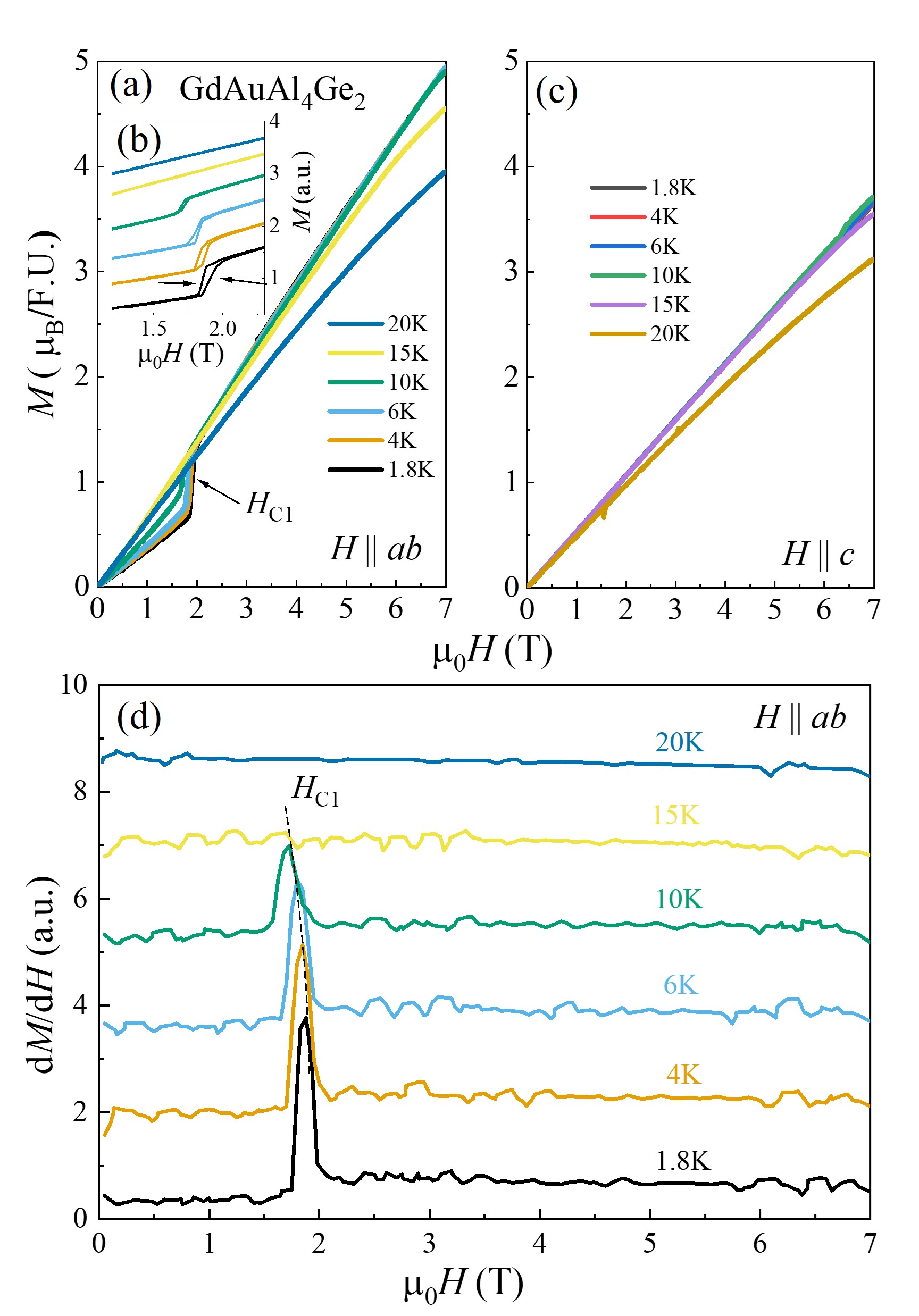}\caption{\label{fig3} (a) Isothermal magnetic field dependent magnetization $M(H)$ for GdAuAl$_4$Ge$_2$ with fields applied in the $ab$ plane. (b) Zoom of the region near the metamagnetic phase transition at $H_{\rm{C1}}$. Data are shifted vertically for clarity. (c) $M(H)$ for fields along the $c$-axis. (d) The derivative of the magnetization $\frac{\partial M}{\partial H}$ for $H$ $\parallel$ $ab$ at various temperature. Data have been shifted vertically for clarity and the dashed line is a guide to the eye.}
\end{figure}

Isothermal magnetization $M(H)$ measurements are shown in Fig.~\ref{fig3}. For $H$ $\|$ $c$ at $T$ = 1.8 K, $M(H)$  increases linearly with applied field and does not saturate by 7 T. For $H$ $\parallel$ $ab$  at $T$ = 1.8 K, $M(H)$ initially exhibits a linear increase in the magnetization, then undergoes an abrupt and hysteretic increases at $H_{\rm{C1}}$ = 1.9 T. Such behavior is typically associated with a first order spin flop transition. $M(H)$ subsequently increases linearly and does not saturate for $\mu_0H$ $<$ 7 T, where it reaches 62 \% of the expected value for Gd$^{3+}$ ($M_{\rm{sat}}$ = 7.94 $\mu_B$/Gd). This transition moves towards lower fields with increasing temperature, where it is seen at 10 K but is no longer observed at 15 K.

\begin{figure}
 \includegraphics[width=\columnwidth]{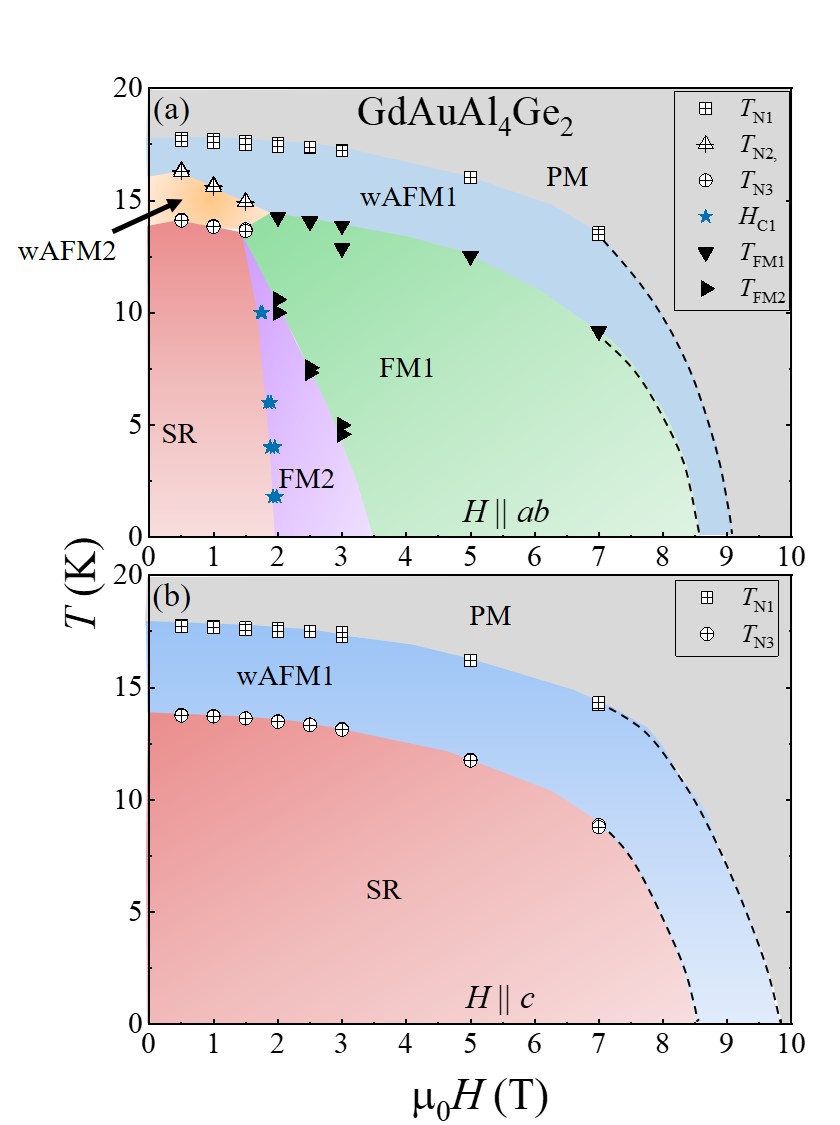}\caption{\label{fig4} Temperature $T$ vs magnetic field $H$ phase diagram for GdAuAl$_4$Ge$_2$ constructed from the magnetic susceptibility $\chi(T)$ and isothermal magnetization $M(H)$. The zero field phase transitions agree with results from heat capacity $C(T)$, and electrical resistivity $\rho(T)$ data, described below. The various regions wAFM1, wAFM2, SR, FM1, and FM2 are described in the text.}
\end{figure}

These results are used to construct the phase diagrams shown in Fig~\ref{fig4}, where a rich family of ordered phases is seen for $H$ $\parallel$ $ab$. Starting from the paramagnetic state at low $H$, we find that the transition at $T_{\rm{N1}}$ weakly reduces $\chi$ upon entering an antiferromagnetic-like region (wAFM1). This is rapidly replaced by another transition that further reduces $\chi$ upon entering another antiferromagnetic-like region (wAFM2). Finally, there is a strong reduction in $\chi$  when the system undergoes a first order phase transition into the spin reoriented ground state (SR). The rapid progression of phases, and the reductions in $\chi$, suggests that the phases are characterized by progressively more anti-aligned spin configurations which may feature complicated wave vectors. Above 1.5 T there is an abrupt change in the ordered state behavior, where the transition into the wAFM state continues to appear as a weak reduction from the paramagnetic behavior, but the low field reductions in $\chi$ are replaced by ferromagnetically spin polarized states labeled FM1 and FM2. These boundaries are suppressed with increasing field, and are extrapoled to collapse towards zero temperature near 8-10 T. The low temperature boundary between SR and FM1 is clearly seen in the field dependent magnetization, where it appears as a hysteretic first order metamagnetic step. These results are contrasted with what is seen when fields are applied along the $c$ axis, where the phase boundaries are gradually suppressed with $H$.
\begin{figure}
  \includegraphics[width=\columnwidth]{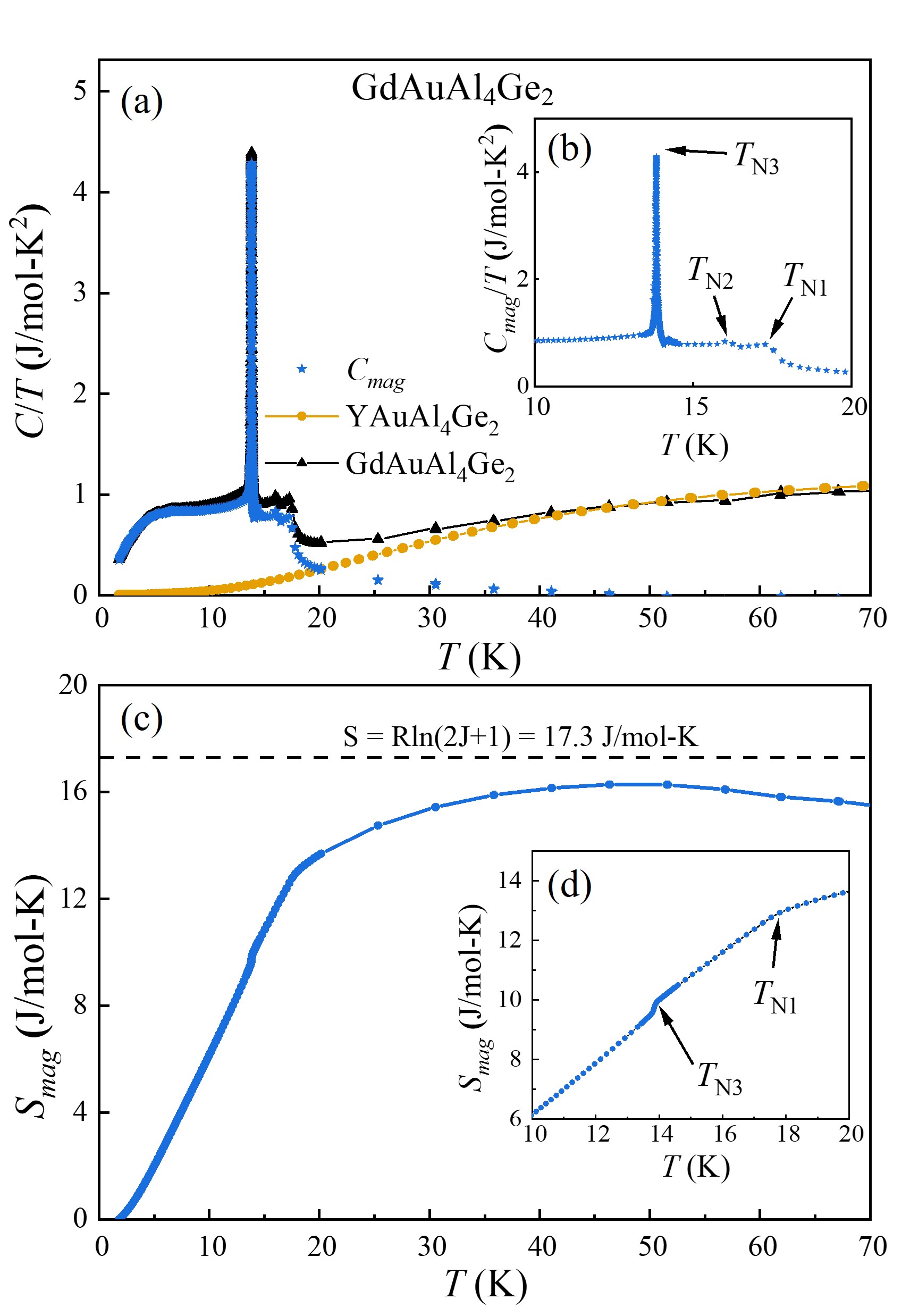}\caption{\label{fig5} (a) The heat capacity divided by temperature $C/T$ vs. $T$ for GdAuAl$_4$Ge$_2$ and YAuAl$_4$Ge$_2$. $C_{\rm{mag}} / T$ is calculated as described in the text. (b) Zoom of $C_{\rm{mag}} / T$ near the transition temperatures. (c) Magnetic entropy entropy $S_{\rm{mag}}$ vs. $T$, which is obtained from the heat capacity data as described in the text. The dotted line represents the calculated entropy for the full $J$ $=$ 7/2 Hund's rule multiplet. (d) Zoom of $S_{\rm{mag}}$ in the region near the magnetic ordering.}
\end{figure}
Heat capacity divided by temperature $C$/$T$ data are compared to those of the $J$ = 0 nonmagnetic analogue YAuAl$_4$Ge$_2$ in Fig.~\ref{fig5}a. As expected, there is close agreement between these curves at elevated temperatures, where phonons dominate the heat capacity. However, for $T$ $\lesssim$ 60 K, the Gd curve begins to deviate from the phonon behavior. In order to expose this trend, we plot the isolated magnetic contribution given by $C_{\rm{mag}}$/$T$ $=$ ($C_{\rm{Gd}}$ - $C_{\rm{Y}}$)/$T$, where a long and increasing tail precedes the ordered states. Consistent with $\chi(T)$, three distinct transitions are subsequently seen at $T_{\rm{N1}}$, $T_{\rm{N2}}$, and $T_{\rm{N3}}$. While the first two have typical second order $\lambda$-shape peaks, the feature at $T_{\rm{N3}}$ is sharp and hysteretic which confirms that it is a first-order phase transition. Finally, we determine the magnetic contribution to the entropy $S_{\rm{mag}}$($T$) (Figs.~\ref{fig5} c,d) by integrating $C_{\rm{mag}}$/$T$ for $T$ $>$ 1.8 K. $S_{\rm{mag}}(T$) reaches 12.8 J mol$^{-1}$ K$^{-1}$ at $T_{\rm{N1}}$, which is reduced from the full theoretical magnetic entropy given by $S_{\rm{mag}}$ = $R$ln(2$J$ + 1) = 17.3 J mol$^{-1}$ K$^{-1}$ ($S$ $=$ 7/2, $L$ $=$ 0, and $J$ $=$ 7/2). After this, $S_{\rm{mag}}$ continues to increase until it reaches a saturated value near 16.4 J mol$^{-1}$ K$^{-1}$ for $T$ $\approx$ 40 K. Given that crystal electric field splitting is not expected to influence the behavior of Gd, this suggests that the excess entropy above $T_{\rm{N1}}$ likely originates from magnetic fluctuations of the Gd ions. 
\begin{figure}
   \includegraphics[width=\columnwidth]{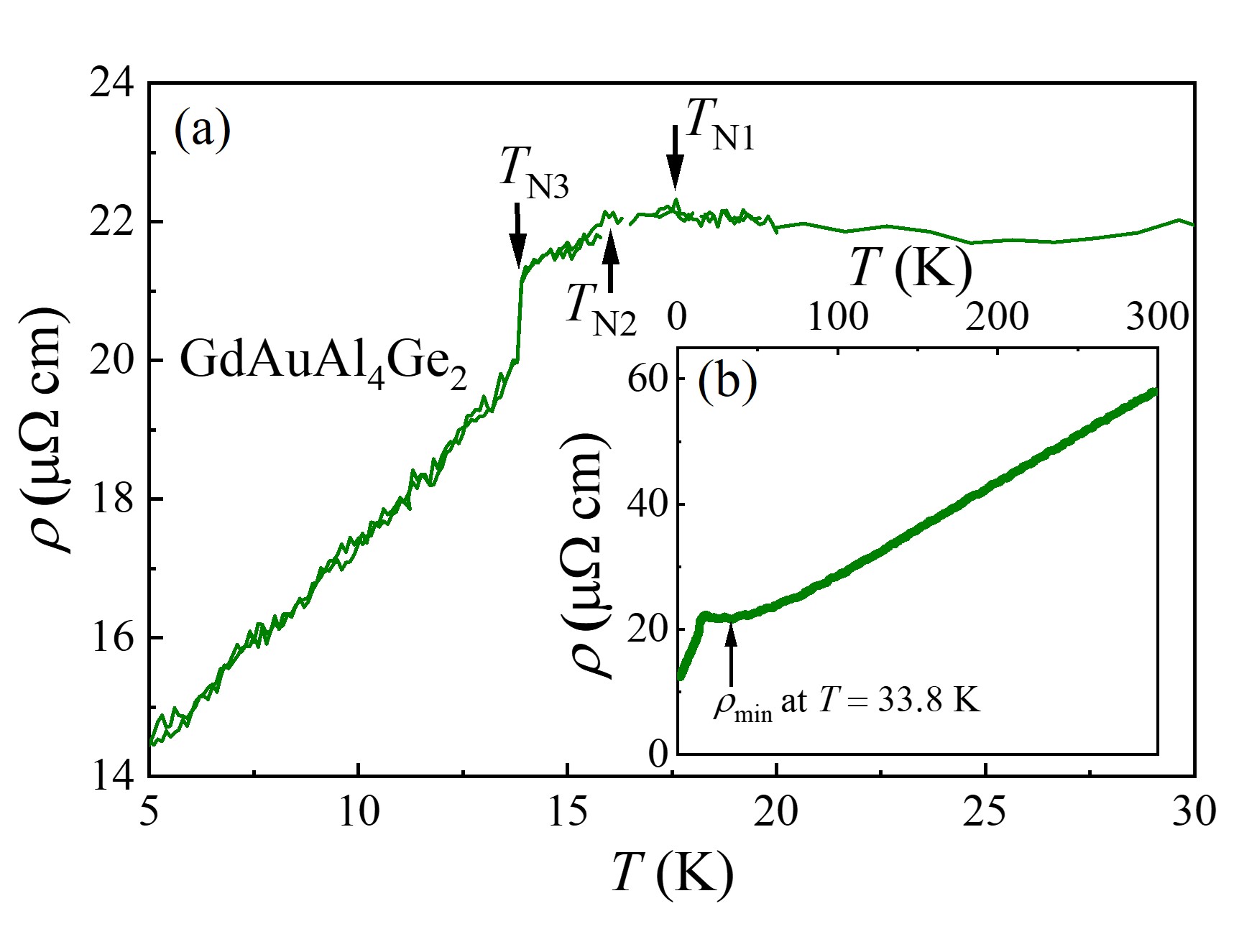}\caption{\label{fig6}  Electrical resistivity $\rho$($T$) at zero magnetic fields with the electrical current applied in an arbitrary direction for GdAuAl$_4$Ge$_2$. The main panel (a) shows the low temperature region near the phase transitions. $T_{\rm{N1}}$ and $T_{\rm{N2}}$ from $\chi$($T$) are marked. The inset figure (b) shows the full temperature range between 1.8 K $<$ $T$ $<$ 300 K. Resistivity minimum at $T$ = 33.8 K is marked with an arrow.}
\end{figure}
Fig.~\ref{fig6} shows the temperature dependent electrical resistivity $\rho$($T$), with the electrical current applied in an arbitrary direction. Metallic behavior is observed from room temperature, where the phonon-electron term is dominant for 50 $<$ $T$ $<$ 300 K. The magnetic ordering is preceded by a weak minimum that is centered near 25 K. This resembles what is seen for Gd$_2$PdSi$_3$, where the minimum in the resistivity is thought to be associated with an interplay between the RKKY interaction and magnetic frustration~\cite{resmin1,Kumar_2020}. $T_{\rm{N1}}$ and $T_{\rm{N2}}$ slightly reduce $\rho$, and there is a strong decrease at $T_{\rm{N3}}$. This shows (i) that spin fluctuations above the ordered state enhance $\rho$ and (ii) that the ordering removes of spin scattering of conduction electrons. We also emphasize that the sharp reduction at $T_{\rm{N3}}$ provides further evidence that this is a first order phase transition.
\subsection{TbAuAl$_4$Ge$_2$}
\begin{figure}
 \includegraphics[width=\columnwidth]{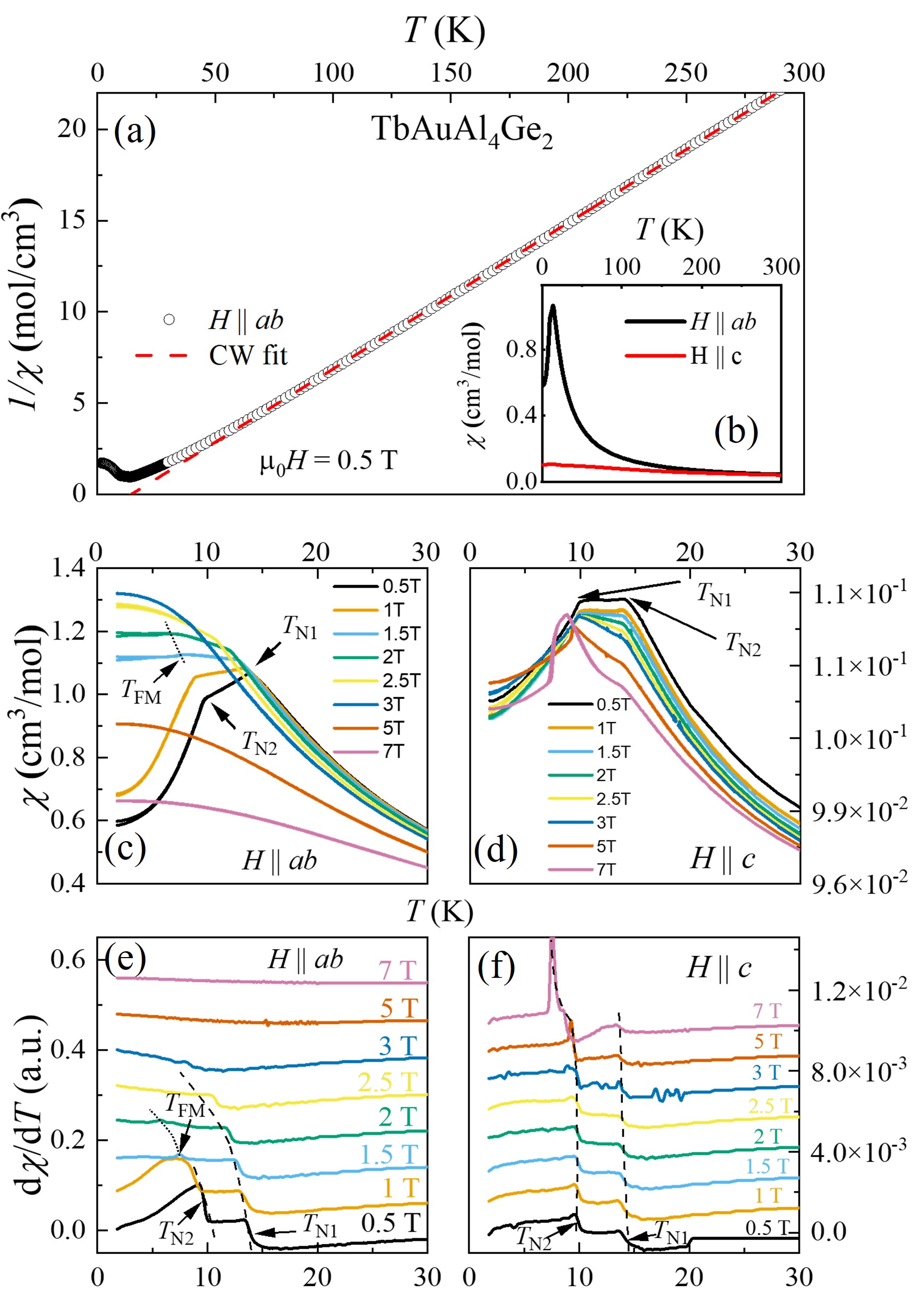}\caption{\label{fig7}(a) The inverse magnetic susceptibility $\chi^{-1} = (M/H)^{-1}$ vs temperature $T$ for TbAuAl$_4$Ge$_2$ collected in a magnetic field $\mu_0H$ $=$ 0.5 T applied parallel to $ab$-plane. The dotted line is a fit to the data using the Curie-Weiss law, as described in the text. (b) $\chi (T)$ for $\mu_0H$ $=$ 0.5 T applied both parallel (black line) to the $ab$-plane and c-axis (red line). (c,d) $\chi (T)$ for 1.8 K $<T<$ 30 K emphasizing the region near the magnetic ordering temperatures $T_{\rm{N1}}$, $T_{\rm{N2}}$, and $T_{\rm{FM}}$ which are defined in the text. (e,f) Derivatives of the magnetic susceptibilities with respect to temperature $\frac{\partial \chi}{\partial T}$. Data have been shifted vertically for clarity, the dashed lines and short dot lines are guided to the eye.}
 \end{figure}

The temperature dependent magnetic susceptibility for TbAuAl$_4$Ge$_2$ is shown in Fig.~\ref{fig7}, where there is strong anisotropy that is associated with the non-zero angular momentum ($L$ = 3). The spins prefer to align in the $ab$-plane, and a Curie-Weiss temperature dependence for 100 K $<$ $T$ $<$ 300 K is observed. Fits to the data yield the parameters $\theta$ = 18.0 K and $\mu_{\rm{eff}}$ = 9.89 $\mu_B$/Tb ($\mu_{\rm{eff}}$ = 9.72 $\mu_B$ for Tb$^{3+}$). For $H\|$ c, the weak $T$ dependence makes it difficult to perform a reliable Curie-Weiss fit. The ordered state behaviors are shown in  Figs.~\ref{fig7}c,d, where antiferromagnetic-like phase transitions at $T_{\rm{N1}}$ = 13.9 K and $T_{\rm{N2}}$ = 9.8 K are seen for $\mu_0H$ $=$ 0.5 T applied in both directions. For $H$ $\parallel$ $ab$, $T_{\rm{N1}}$ and $T_{\rm{N2}}$ are suppressed with increasing $\mu_0H$ up to 1 T, after which larger fields produce a field polarized state with a subphase that appears as a weak reduction of $\chi(T)$ at $T_{\rm{FM}}$.  Figs.~\ref{fig7}e,f shows $\partial \chi/\partial T$, where for $H$ $\parallel$ $ab$ the transitions at $T_{\rm{N1}}$ and $T_{\rm{N2}}$ are seen as steps that are visible up to $\mu_0H$ $=$ 1 T. Above this, $T_{\rm{N1}}$ is suppressed with increasing $H$ until it is no longer observed above 3 T. There is also weak feature within the field polarized state at $T_{\rm{FM}}$ that resembles what is seen for the Gd analogue. Further work is needed to clarify the spin orientation that is associated with it. For $H$ $\parallel$ $c$ the features at $T_{\rm{N1}}$ and $T_{\rm{N2}}$ evolve in a more complicated way. While $T_{\rm{N1}}$ is nearly field independent, $T_{\rm{N2}}$ evolves from being a second-order-like reduction in $\chi$ towards a much sharper decrease, which resembles what is seen for the Gd analogue in low fields.
\begin{figure}
  \includegraphics[width=\columnwidth]{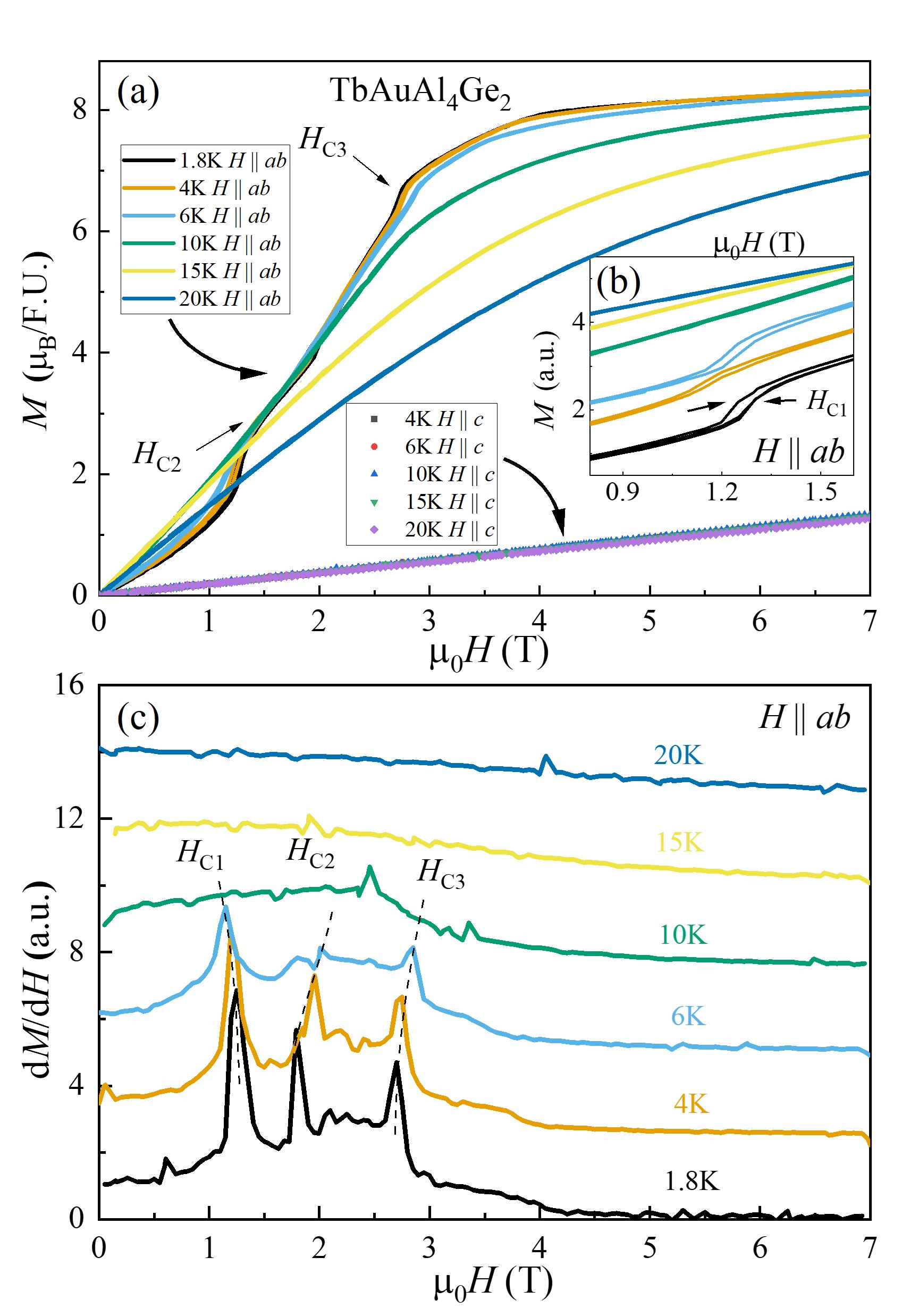}\caption{\label{fig8} (a) Isothermal magnetic field dependent magnetization $M(H)$ for TbAuAl$_4$Ge$_2$ with fields applied in the $ab$ plane and along the $c$-axis. (b) Zoom of $M(H)$ near the metamagnetic phase transitions at $H_{\rm{C1}}$, $H_{\rm{C2}}$, and $H_{\rm{C3}}$. Data are shifted vertically for clarity. (c) The derivative of the magnetization $\frac{\partial M}{\partial H}$ for $H$ $\parallel$ $ab$ at various temperature. The curves have been shifted vertically and the dashed lines are guides to the eye.}
\end{figure}
Isothermal magnetization $M(H)$ measurements are shown in Fig.~\ref{fig8}. For $H$ $\parallel$ $ab$ at $T$ $=$ 1.8 K, the magnetization initially increases linearly and undergoes an abrupt and hysteretic increase at $H_{\rm{C1}}$ $=$ 1.3 T. This is followed by several additional transitions at $H_{\rm{C2}}$ $=$ 1.9 T and $H_{\rm{C3}}$ $=$ 2.7 T, which are revealed in the derivative of the magnetization $\partial$$M$/$\partial$$H$ (Fig.~\ref{fig8}c). The magnetization finally reaches a value of 8.31 $\mu_B$/F.U at 7 T, which is 85\% of the expected saturation value ($M_{\rm{sat}}$ = 9.72 $\mu_B$/Tb). Increasing temperature causes the transitions to broaden and while $H_{\rm{C1}}$ is gradually suppressed, $H_{\rm{C2}}$ and $H_{\rm{C3}}$ move towards larger values. In contrast to this, $M(H)$ for $H$ $\parallel$ $c$ increases linearly with $H$ and shows no evidence for any transitions. 
\begin{figure}
 \includegraphics[width=\columnwidth]{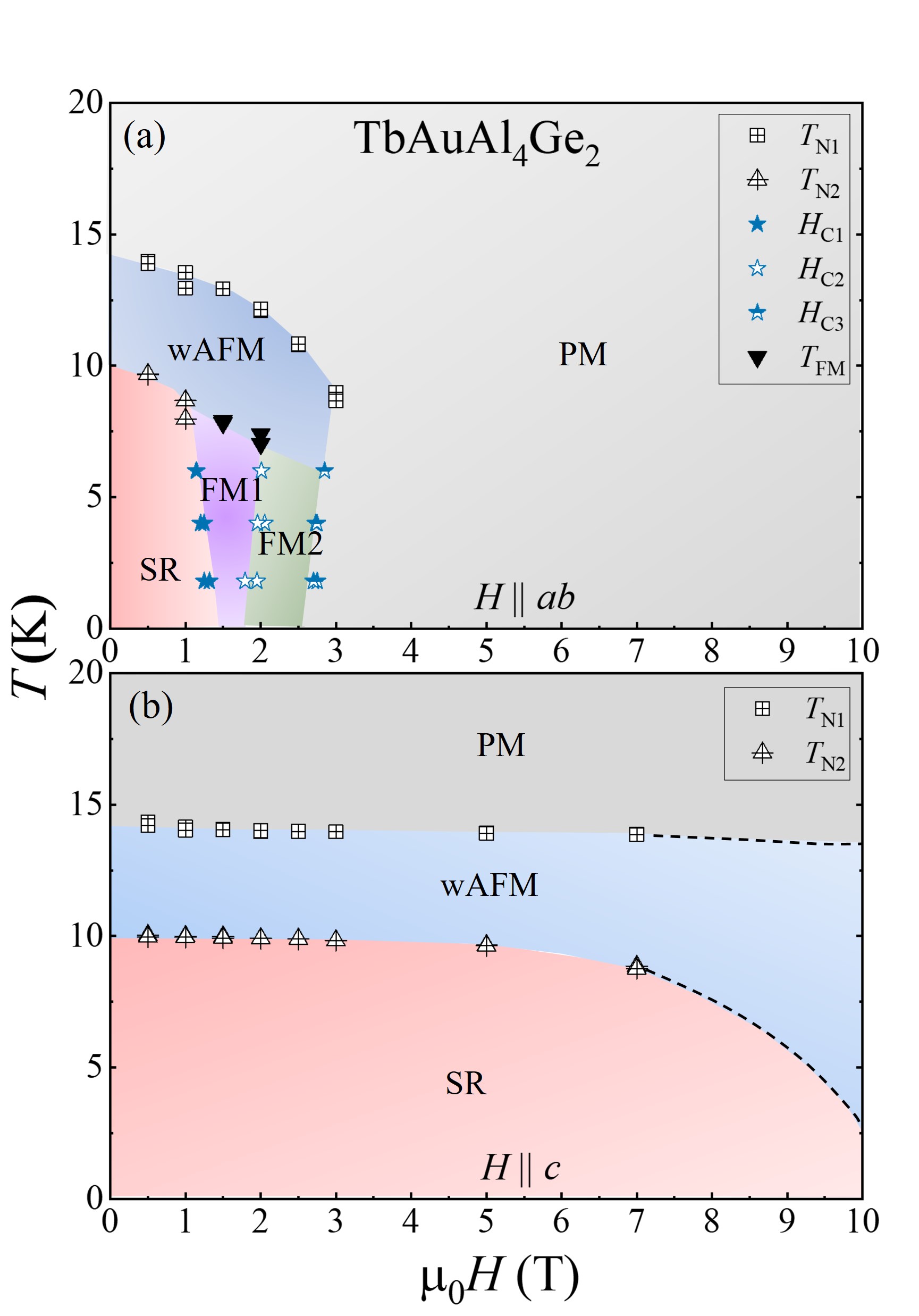}\caption{\label{fig9} Temperature $T$ vs magnetic field $H$ phase diagram for TbAuAl$_4$Ge$_2$ constructed from the magnetic susceptibility $\chi(T)$ and isothermal magnetization $M(H)$. The zero field phase transition agree with results from heat capacity $C(T)$, and electrical resistivity $\rho(T)$ data, described below. The various regions wAFM1, wAFM2, SR, FM1, and FM2 are described in the text.}
\end{figure}
The resulting phase diagrams are shown in Fig~\ref{fig9}, where there are noteworthy similarities to what is seen for GdAuAl$_4$Ge$_2$. For $H$ $\parallel$ $ab$ and at low $H$, the system undergoes a second order transition into a state that reduces $\chi$ below the extrapolated paramagnetic curve (wAFM1). This is rapidly followed by another second order transition that further reduces $\chi$ (SR). Again, this suggests that each of these transitions represents progressively more antialigned spin configurations with complicated wave vectors. Above $\mu_0H$ $\approx$ 1.5 T there is an abrupt change in the ordered state behavior, where the transition into the wAFM state continues to appear as a weak reduction from the paramagnetic behavior, but the low field reductions in $\chi$ are replaced by spin polarized states labeled FM1 and FM2. These boundaries are suppressed with increasing field, but unlike what is seen for GdAuAl$_4$Ge$_2$, are sharply truncated and collapse towards zero temperature near 2.5-3 T. The low temperature boundaries between SR,  FM1, and FM2 are clearly seen in the field dependent magnetization, where the boundary between SR and FM1 appears as a hysteretic first order metamagnetic step, and the subsequent boundaries are broader increases in $M$. The $T-H$ phase diagram is much simpler when fields are applied along the $c$ axis, where the phase boundaries are gradually suppressed with $H$.
\begin{figure}[ht]
  \includegraphics[width=\columnwidth]{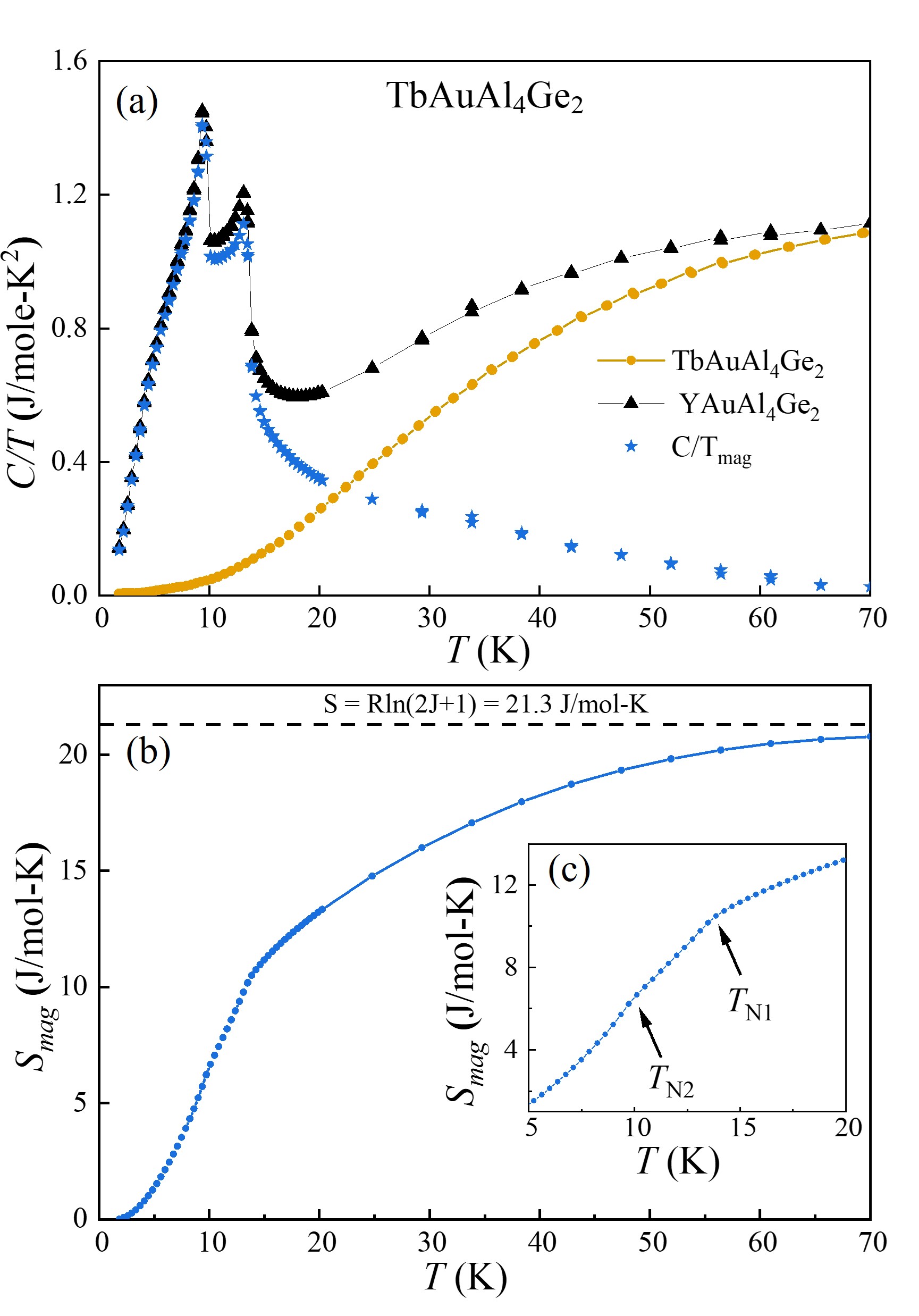}\caption{\label{fig10}(a) The heat capacity divided by temperature $C/T$ vs. $T$ for TbAuAl$_4$Ge$_2$ and YAuAl$_4$Ge$_2$. $C_{\rm{mag}} / T$ is calculated as described in the text. (b) Magnetic entropy entropy $S_{\rm{mag}}$ vs. $T$, which is obtained from the heat capacity data as described in the text. The dotted line represents the calculated entropy for the full $J$ $=$ 6 Hund's rule multiplet. (c) Zoom of $S_{\rm{mag}}$ in the region near the magnetic ordering temperatures.}
\end{figure}
The heat capacity divided by temperature $C$/$T$ data are compared to those for YAuAl$_4$Ge$_2$ in Fig.~\ref{fig10}. As for the Gd analogue, at elevated temperatures  there is close agreement between the data sets but for $T$ $\lesssim$ 70 K, the data deviate from what is seen for the Y compound. The isolated magnetic contribution $C_{\rm{mag}}$/$T$ $=$ ($C_{\rm{Tb}}$ - $C_{\rm{Y}}$)/$T$ reveals that a long and increasing tail precedes the ordered states and that ordering is seen at $T_{\rm{N1}}$ = 13.9 K, $T_{\rm{N2}}$ = 9.8 K as second-order peaks. The magnetic entropy $S_{\rm{mag}}$($T$) is shown in Fig.~\ref{fig10}c, where $S_{\rm{mag}}(T$) reaches 13.9 J mol$^{-1}$ K$^{-1}$ at $T_{\rm{N1}}$. Again this value is reduced from the expected value $S_{\rm{mag}}$ = $R$ln(2$J$ + 1) = 21.3 J mol$^{-1}$ K$^{-1}$ ($S$ $=$ 3, $L$ $=$ 3, and $J$ $=$ 6). After this it continues to increase until reaching a saturation value of 20.7 J mol$^{-1}$ near $T$ $\approx$ 70 K. This provides evidence that (i) the crystal electric field split levels are fully populated around this temperature and (ii) magnetic fluctuations may extend well above the ordered state, although the influences of these factors are not distinguishable. 
\begin{figure}
   \includegraphics[width=\columnwidth]{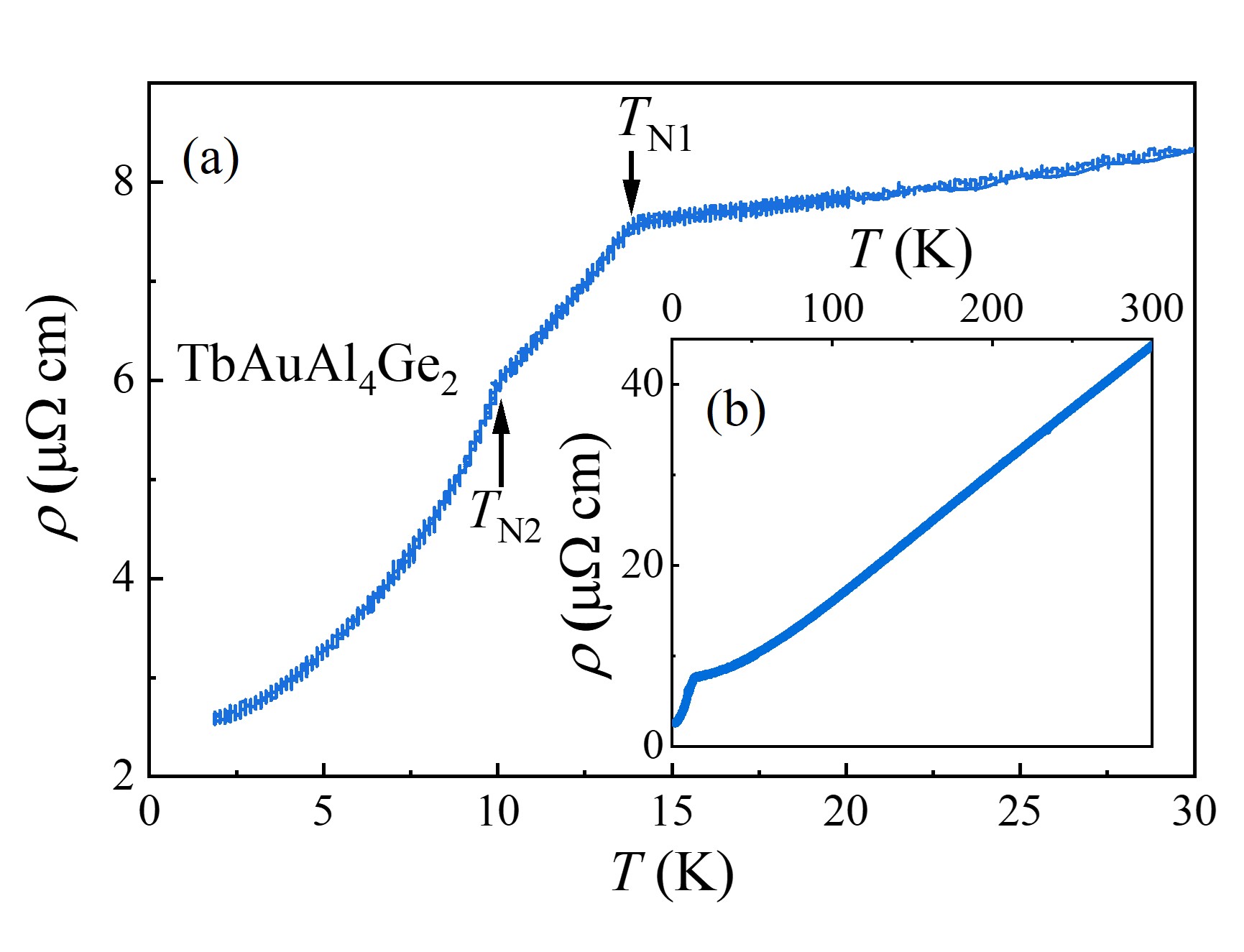}\caption{\label{fig11} Electrical resistivity $\rho$($T$) at zero magnetic field with the electrical current applied in an arbitrary direction for TbAuAl$_4$Ge$_2$. The main panel show the low temperature region near the phase transitions and the inset shows the full temperature range between 1.8 K $<$ $T$ $<$ 300 K.}
\end{figure}
Fig.~\ref{fig11} shows the temperature dependence of the electrical resistivity $\rho$($T$) for TbAuAl$_4$Ge$_2$ with the electrical current applied in an arbitrary direction. Metallic behavior is observed from room temperature, where the phonon-electron term is dominant for 50 $<$ $T$ $<$ 300 K. It is noteworthy that the minimum that is seen for the Gd analogue does not appear here. Finally, the rapid decrease in $\rho(T)$ coincides with the onset of the magnetic ordering at $T_{\rm{N1}}$ and indicates the significant removal of spin scattering of conduction electrons.

\section{\label{sec:Discussion}Discussion}

Taken together, these measurements show that both GdAuAl$_4$Ge$_2$ and TbAuAl$_4$Ge$_2$ exhibit complex magnetic phenomena with noteworthy similarities. Evidence for this includes (i) magnetic entropy that extends to temperatures well above the ordered states, (ii) a sharp drop of resistivity as the magnetic order sets in, and (iii) the occurrence of multiple magnetically ordered states that are delicately tuned by magnetic fields. These features are shared by other other magnetically frustrated materials that have nontrivial spin textures (e.g., Gd$_2$PdSi$_3$, GdRu$_2$Si$_2$, and Gd$_3$Ru$_4$Al$_{12}$\cite{Skyrmion.centrosymmetric.condition.Nature.Communication.2015,skyrmion.Centrosymmetric.Magnets.PRL.2018,Gd3Ru4Al12.skyromion.Nature.Communication.2019,GdRu2Si2.Nature.2020,Gd2PdSi3.skyrmion.science.2019}) and are currently motivating investigations of a variety of related materials (e.g., $Ln_2$RhSi$_3$ $Ln$ = Gd, Tb, Dy~\cite{Kumar_2020}). However, it is also well known that Gd and Tb based metals exhibit a wide variety of complex magnetic states with no obvious connection to skyrmion states~\cite{GdHC,handbook}.  Based on this, we propose that further studies that target the magnetic structure such as neutron scattering, Lorentz tunnelling electron microscopy, and magnetic force microscopy will be of interest to clarify the behavior of the spins in these compounds. Also needed are efforts to investigate whether these compounds exhibit novel electrical transport properties such as a topological Hall effect.

Irrespective of whether any of these states exhibit nontrivial electronic behaviors such as the topological Hall effect, it will also be of interest to understand the various magnetically ordered phases that are present, the factors that drive differences or similarities between the Gd and Tb examples, and their interactions with other degrees of freedom. For example, it is currently unclear why the Gd case exhibits a cascade of transitions between the wAFM1, wAFM2, and SR states, along with a first order phase transition, while the Tb example does not. We speculate that this relates to differences in the magnetic interactions that are associated with the isotropic pure spin Gd ions and the anisotropic Tb ions, but more work is needed to verify this. It is also noteworthy that while GdAuAl$_4$Ge$_2$ exhibits a resistivity minimum that precedes the ordering, TbAuAl$_4$Ge$_2$ does not. This implies that variation of the lanthanide ions will likely be useful to optimize both magnetic frustration and the conditions for the formation of a possible skyrmion lattice.

\section{\label{sec:Conclusions}Conclusions}

In summary, we have produced the series of compounds $Ln$AuAl$_4$Ge$_2$ ($Ln$ = Y, Pr, Nd, Sm, Gd, Tb, Dy, Ho, Er, and Tm) and have focused on the bulk magnetic properties of the Gd and Tb variants. Temperature and magnetic field dependent magnetization, heat capacity, and electrical resistivity measurements show that both of these compounds exhibit several magnetically ordered states at low temperatures, with evidence for magnetic fluctuations extending into the paramagnetic temperature region. Applied magnetic fields produce several distinct magnetically ordered regions, where there are many similarities between the two compounds. This is surprising given their different orbital quantum numbers, and suggests that the features of the crystalline lattice or the Fermi surface topography are dominant factors in determining the ground state behavior. Thus, the family of materials $Ln$AuAl$_2$Ge$_2$ emerges as a reservoir for novel metallic magnetism and invites investigations into them to search for nontrivial spin structures and novel electronic behavior such as the topological Hall effect.

\section{\label{sec:results}Acknowledgements}
RB, KF, and OO were supported by the National Science Foundation through NSF DMR-1904361. Work at the University of Colorado Boulder was supported by Award No. DE-SC0021377 of the U.S. Department of Energy, Basic Energy Sciences, Materials Sciences and Engineering Division. The National High Magnetic Field Laboratory is supported by the National Science Foundation through NSF DMR-1644779 and the
State of Florida. 

\bibliography{Reference}
\end{document}